\documentstyle [12pt] {article}

\catcode`@=12
\begin{document}
\thispagestyle{empty}
\begin{flushright} UCRHEP-T271\\February 2000\
\end{flushright}
\vskip 0.5in
\begin{center}
{\Large \bf Neutrino Exotica in the Skew E$_6$ Left-Right Model\\}
\vskip 2.0in
{\bf Ernest Ma\\}
\vskip 0.3in
{\sl Physics Department, Univeristy of California, Riverside, 
CA 92521, USA\\}
\end{center}
\vskip 1.8in
\begin{abstract}\
With the particle content of the {\bf 27} representation of E$_6$, a skew 
left-right supersymmetric gauge model was proposed many years ago, with a 
variety of interesting phenomenological implications.  The neutrino sector 
of this model offers a natural framework for obtaining small Majorana 
masses for $\nu_e$, $\nu_\mu$, and $\nu_\tau$, with the added bonus of 
accommodating 2 light sterile neutrinos.
\end{abstract}
\newpage
\baselineskip 24pt

With the advent of superstring theory\cite{string}, it was recognized early 
on\cite{heterotic} that the gauge symmetry E$_6$ may be relevant for 
discussing low-energy particle physics phenomenology\cite{hewriz}.  There are 
two ideas: (1) the particle content of the Minimal Supersymmetric Standard 
Model (MSSM) may be extended to include all particles contained in the 
fundamental {\bf 27} representation of E$_6$; and (2) the standard-model 
gauge group may be extended as well.  The most actively pursued such 
approach\cite{cvelan,chohag} is to add an extra U(1).

A very different and unique alternative was proposed many years ago\cite{ma}, 
which considers instead an unconventional $SU(3)_C \times SU(2)_L \times 
SU(2)_R \times U(1)$ decomposition of the {\bf 27} representation of E$_6$, 
resulting in a variety of interesting phenomenological implications. 
Among these are the natural absence of flavor-changing neutral currents at 
tree level\cite{ma} despite the presence of $SU(2)_R$, the possibility of 
breaking $SU(2)_R$ at or below the TeV scale with only Higgs doublets and 
bidoublets\cite{bahema} without conflicting with present phenomenology, and 
the appearance of an effective two-doublet Higgs sector different\cite{mang1} 
from that of the MSSM, with the interesting (and currently very relevant) 
property that the tree-level upper bound of the lightest neutral Higgs-boson 
mass is raised from $M_Z$ in the MSSM to $\sqrt 2 M_W$ in this case.

Two possible deviations from the standard model have recently been observed. 
One is a new determination of the weak charge of atomic cesium\cite{bewi}.  
The other is a new analysis of the hadronic peak cross section at the $Z$ 
resonance\cite{mnsw}.  Based on these data, it has now been 
shown\cite{erla} that the model of Ref.\cite{ma} is in fact the most favored 
of all known gauge extensions of the standard model.

This paper deals with another aspect of this remarkable model, i.e.~that of 
its neutrinos.  In the original proposal\cite{ma}, neutrino masses were 
{\it assumed} to be zero for simplicity.  [Recall that in 1986, neutrino 
oscillations were not clearly established.]  However, such is not an essential 
feature of this model.  It is in fact more natural that $\nu_e$, $\nu_\mu$, 
and $\nu_\tau$ acquire small Majorana masses through the usual sesaw 
mechanism\cite{seesaw}, and that 2 light sterile neutrinos are accommodated 
in this model, resulting in a number of exotic phenomena which may be tested 
in future experiments.

The skew E$_6$ left-right model is based on the observation that there are 
two ways of identifying the standard-model content of the {\bf 27} 
representation of E$_6$.  Written in its [$SO(10)$, $SU(5)$] decomposition, 
we have
\begin{eqnarray}
{\bf 27} &=& (16,5^*) + (16,10) + (16,1) \nonumber \\ 
&+& (10, 5^*) + (10,5) + (1,1).
\end{eqnarray}
The usual assumption is that the standard-model particles are contained 
in the $(16,5^*)$ and $(16,10)$ multiplets.  On the other hand, if we 
switch $(16,5^*)$ with $(10,5^*)$ and $(16,1)$ with $(1,1)$, the 
standard model remains the same.  The difference between the 2 options 
only appears if the gauge group is extended.  In particular, a very 
different and unique model emerges if the gauge group becomes $SU(3)_C 
\times SU(2)_L \times SU(2)_R \times U(1)$.  In this scenario, the particle 
assignments are as follows.
\begin{eqnarray}
&& (u,d)_L \sim (3,2,1,1/6), ~~~ d^c_L \sim (3^*,1,1,1/3), \\ 
&& (h^c,u^c)_L \sim (3^*,1,2,-1/6), ~~~ h_L \sim (3,1,1,-1/3), \\ 
&& \left( \begin{array} {c@{\quad}c} \nu & E^c \\ e & \psi^0 
\end{array} \right)_L \sim (1,2,2,0), ~~~ (e^c,S)_L \sim (1,1,2,1/2),  \\ 
&& (\xi^0,E)_L \sim (1,2,1,-1/2), ~~~ N_L \sim (1,1,1,0),
\end{eqnarray}
where the convention is that all fields are considered left-handed.

The notion of $R$ parity is an important ingredient of this construction. 
The usual quarks and leptons, i.e.~$u$, $d$, $u^c$, $d^c$, $\nu$, $e$, and 
$e^c$, with the addition of $N$, have $R = +1$ and their scalar 
supersymmetric partners have $R = -1$ as in the MSSM.  The other fermions, 
i.e.~$h$, $h^c$, $E$, $E^c$, $\psi^0$, $\xi^0$, and $S$ have $R = -1$ and 
their scalar supersymmetric partners have $R = +1$.  Furthermore, all gauge 
bosons have $R = +1$ and gauge fermions have $R = -1$, {\it except} $W_R^\pm$ 
which has $R = -1$ and $\tilde W_R^\pm$ which has $R = +1$.  This unusual 
feature is the origin of many desirable and interesting 
properties\cite{hewriz} of this model which sets it apart from all other 
gauge extensions of the standard model.

Consider the $R = +1$ neutral fermion sector, i.e.~the usual neutrinos 
$\nu_e$, $\nu_\mu$, $\nu_\tau$, and the 3 $N$'s.  They are linked by the 
Yukawa terms $\nu_i N_j \tilde \psi^0_k$, where one linear combination of 
$\tilde \psi^0_k$ may be identified with the usual Higgs scalar $h_2^0$ 
which acquires the vacuum expectation value $v_2$.  Furthermore, since 
$N_j$ transforms {\it trivially} under $SU(3)_C \times SU(2)_L \times SU(2)_R 
\times U(1)$, it is allowed to have a nonzero Majorana mass which is 
presumably large\cite{nasa}.  [In U(1) extensions of the MSSM within the 
context of E$_6$, the requirement that $N$ transform trivially under $SU(3)_C 
\times SU(2)_L \times U(1)_Y \times U(1)$ uniquely determines it\cite{make} 
to be a particular linear combination\cite{hewriz} of $U(1)_\psi$ and 
$U(1)_\chi$ with mixing angle $\alpha = - \tan^{-1} \sqrt {1/15}$, where 
$Q_\alpha = Q_\psi \cos \alpha - Q_\chi \sin \alpha$.  This is referred to 
as $U(1)_N$ or $U(1)_\nu$\cite{ibma}.  However, with two U(1) gauge factors, 
kinetic mixing\cite{holdom} must be considered, a complication which is 
absent in a left-right model.]  The resulting $6 \times 6$ neutrino mass 
matrix is
\begin{equation}
{\cal M}_\nu = \left( \begin{array} {c@{\quad}c} 0 & m_D \\ m_D^T & m_N 
\end{array} \right),
\end{equation}
where $m_D$ and $m_N$ are themselves $3 \times 3$ matrices.  Thus the usual 
neutrinos acquire small Majorana masses through the canonical seesaw 
mechanism\cite{seesaw} without any problem.  In other gauge extensions, 
this mechanism is often not available\cite{erla}.

The $SU(2)_R \times U(1)$ of this model breaks down to the standard-model 
$U(1)_Y$ through the vacuum expectation value $v_3$ of a linear combination 
of the $\tilde S$'s.  Let us define that to be $\tilde S_3$.  Because of the 
allowed Yukawa terms linking $h h^c$, $E E^c$, and $\xi^0 \psi^0$ to 
$\tilde S_3$, these exotic fermions have masses proportional to $v_3$. 
However, only 1 of the 3 $S$'s gets a mass at this stage, i.e.~$S_3$, as 
it is linked to a particular linear combination of the two neutral gauge 
fermions corresponding to $SU(2)_R$ and $U(1)$ through $\tilde S_3$. 

Electroweak symmetry breaking proceeds as in the MSSM, with $\tilde \xi_3$ 
identified as $h_1^0$ and $\tilde \psi_3$ as $h_2^0$, having vacuum 
expectation values $v_1$ and $v_2$ respectively.  Now $S_{1,2}$ are no 
longer massless and if they are light, they could well be called sterile 
neutrinos\cite{sterile}. 

Consider now the $R = -1$ neutral fermion sector, i.e.~$\xi^0_i$, $\psi^0_i$, 
$S_i$, and the 3 gauge fermions $\tilde W_L^0$, $\tilde W_R^0$, and $\tilde B$ 
corresponding to $SU(2)_L$, $SU(2)_R$, and $U(1)$ respectively.  They are 
linked by the
\begin{equation}
f_{ijk} (\xi^0_i e_j e^c_k - E_i \nu_j e^c_k - \xi^0_i \psi^0_j S_k + 
E_i E^c_j S_k)
\end{equation}
terms of the superpotential as well as the gauge interaction terms together 
with the soft supersymmetry-breaking Majorana mass terms $m_L$, $m_R$, $m_B$ 
of the gauge fermions.  The resulting $12 \times 12$ mass matrix is
\begin{equation}
{\cal M} = \left[ \begin{array} {c@{\quad}c@{\quad}c@{\quad}c} 0 & -m_{EE^c} 
& -f_{i3j} v_2 & m_1 \\ -m_{EE^c}^T & 0 & -m_{ee^c} & m_2 \\ -f_{j3i} v_2 & 
-m_{ee^c}^T & 0 & m_3 \\ m_1^T & m_2^T & m_3^T & \tilde m \end{array} 
\right],
\end{equation}
where
\begin{eqnarray}
m_1 = {v_1 \over \sqrt 2} \left( \begin{array} {c@{\quad}c@{\quad}c} 0 & 0 & 
0 \\ 0 & 0 & 0 \\ g_L & 0 & -g_B \end{array} \right), &~& 
m_2 = {v_2 \over \sqrt 2} \left( \begin{array} {c@{\quad}c@{\quad}c} 0 & 0 & 
0 \\ 0 & 0 & 0 \\ -g_L & g_R & 0 \end{array} \right), \\
m_3 = {v_3 \over \sqrt 2} \left( \begin{array} {c@{\quad}c@{\quad}c} 0 & 0 & 
0 \\ 0 & 0 & 0 \\ 0 & -g_R & g_B \end{array} \right), &~&
\tilde m = \left( \begin{array} {c@{\quad}c@{\quad}c} m_L & 0 & 0 \\ 
0 & m_R & 0 \\ 0 & 0 & m_B \end{array} \right).
\end{eqnarray}
The gauge couplings $g_L$, $g_R$, and $g_B$ are related to the electromagnetic 
coupling $e$ by
\begin{equation}
{1 \over e^2} = {1 \over g_L^2} + {1 \over g_R^2} + {1 \over g_B^2}.
\end{equation}
Assuming that $g_L = g_R$, we then have
\begin{equation}
g_L^2 = g_R^2 = {e^2 \over \sin^2 \theta_W}, ~~~ g_B^2 = {e^2 \over 1 - 2 
\sin^2 \theta_W},
\end{equation}
where $\theta_W$ is the usual electroweak mixing angle.

It is clear from the above that the $10 \times 10$ mass submatrix spanning 
$\xi^0_i$, $\psi^0_i$, $S_3$, and the 3 gauge fermions will have 8 eigenvalues 
of order $v_3$ and 2 eigenvalues of order $m_{L,R,B}$.  The $2 \times 2$ 
effective mass matrix spanning $S_{1,2}$ is then given by a generalized 
seesaw formula, i.e.
\begin{equation}
({\cal M}_S)_{ij} = \sum^3_{k=1} \sum^3_{l=1} {2 f_{k3i} v_2 (m_{ee^c})_{lj} 
\over (m_{EE^c})_{kl}}.
\end{equation}
If $f_{k3i}$ are small enough, say less than $m_{ee^c}/v_2$, then $S_{1,2}$ 
may indeed be light enough to be considered as sterile neutrinos.  Note that 
under the standard $SU(3)_C \times SU(2)_L \times U(1)_Y$ gauge group, 
$S \sim (1,1,0)$ is indeed a singlet.

We now have 5 light neutrinos, the usual ones $\nu_e$, $\nu_\mu$, $\nu_\tau$ 
with $R = +1$, and the sterile ones $S_{1,2}$ with $R = -1$.  Since $R$ 
parity is still strictly conserved, they do not mix.  Hence $S_{1,2}$ would 
not be a factor in considering the phenomenology of neutrino oscillations. 
However, as a discrete symmetry, $R$ parity may be broken by {\it soft} 
terms\cite{whepp2} without affecting other essential properties of the 
unbroken theory.  Remarkably, exactly such a soft term is allowed by the 
$SU(3)_C \times SU(2)_L \times SU(2)_R \times U(1)$ gauge symmetry of this 
model, namely
\begin{equation}
m'_{ij} (\nu_i \psi^0_j - e_i E^c_j).
\end{equation}
Hence the $3 \times 2$ matrix linking $\nu_e$, $\nu_\mu$, $\nu_\tau$ with 
$S_{1,2}$ is given by
\begin{equation}
({\cal M}_{\nu S})_{ij} = \sum^3_{k=1} \sum^3_{l=1} {m'_{il} f_{k3j} v_2 
\over (m_{EE^c})_{kl}},
\end{equation}
and we have a general theoretical framework for considering 3 active and 2 
sterile neutrinos.  Of course, further assumptions would be necessary to 
obtain a desirable pattern to explain the present observations of neutrino 
oscillations\cite{atm,sol,lab}.

Given that $S_{1,2}$ are light, the fact that they are connected to $e^c_i$ 
through $W_R^\pm$ means that there are many modifications to the 
phenomenology of lepton weak interactions.  One possibility is that $v_3$ is 
very large, then all such deviations are negligible.  On the other hand, the 
analysis of Ref.\cite{erla} shows that it is possible to have $v_3$ of order 
1 TeV or less.  This would allow experiments in the near future to observe 
a number of exotic phenomena as discussed below.

Whereas the neutral gauge bosons of this model are flavor-diagonal in their 
couplings, the scalar bosons are not.  Hence there will be some 
flavor-changing interactions, most of which are suppressed by Yukawa 
couplings.  However, there is one important exception, i.e.~those terms 
proportional to the top-quark mass.  They appear in the superpotential as 
the following gauge-invariant combination:
\begin{equation}
u_i u^c_j \psi^0_k - d_i u^c_j E^c_k - u_i h^c_j e_k + d_i h^c_j \nu_k.
\end{equation}
It was pointed out a long time ago\cite{mang2} that $\tilde h^c$ exchange 
would then contribute significantly to the rare decay $K^+ \to \pi^+ \nu 
\bar \nu$.  As it happens, the first measurement\cite{adler} of this 
branching fraction, i.e.
\begin{equation}
B (K^+ \to \pi^+ \nu \bar \nu) = 4.2 \begin{array} {c} +9.7 \\ -3.5 
\end{array} \times 10^{-10},
\end{equation}
is in fact somewhat larger than the standard-model expectation\cite{bubu}, 
$(0.82 \pm 0.32) \times 10^{-10}$.  If that is the correct interpretation, 
then using the results of Ref.\cite{mang2}, another prediction of this model 
is
\begin{equation}
{B(b \to s \nu \bar \nu) \over B(K^+ \to \pi^+ \nu \bar \nu)} \simeq 2.4 
{|V_{tb} V_{us}|^2 \over |V_{cb} V_{td}|^2},
\end{equation}
which is of order $10^5$.  Hence the branching fraction of $b \to s \nu \bar 
\nu$ should be about $10^{-5}$, which is several orders of 
magnitude above the standard-model expectation.

Whereas the 2 light sterile neutrinos $S_{1,2}$ do not have standard-model 
interactions, they do transform nontrivially under $SU(2)_R \times U(1)$. 
Hence they must interact with the new heavy gauge bosons $W_R^\pm$ and
\begin{equation}
Z' = -\left({1-2x \over 1-x}\right)^{1 \over 2} W_R^0 + \left({x \over 1-x}
\right)^{1 \over 2} B,
\end{equation}
where $x \equiv \sin^2 \theta_W$.  Consequently, there are several essential 
features of this model involving $S_{1,2}$.

(A) The fundamental weak decay $\mu^- \to e^- \nu_\mu \bar \nu_e$ is now 
supplemented with $\mu^- \to e^- S_i \bar S_j$ from $W_R^\pm$ exchange. 
The latter is constrained by present data\cite{pdg} through the limit 
\begin{equation}
|g^V_{RR}| =  \left( 1 - U^2_{\mu 3} \right)^{1 \over 2} \left( 1 - U^2_{e3} 
\right)^{1 \over 2} (m_{W_L}^2 /m_{W_R}^2) < 0.033,
\end{equation}
where $S_e = U_{e1} S_1 + U_{e2} S_2 + U_{e3} S_3$, etc. and the matrix $U$ 
has been assumed real for simplicity.

(B) The flavor-changing decay $\mu \to e e e$ gets tree-level contributions 
from Eq.~(7) through $\tilde \xi_i^0$ exchange.  However, they may be very 
small because the $f_{ijk}$'s may be chosen arbitrarily to supress any 
such effect in this case.  In contrast, there is an unavoidable one-loop 
contribution from the exchange of $W_R^\pm$ and $S_i$, which is the analog 
of the standard-model case of $W_L^\pm$ and $\nu_i$.  Whereas the latter 
is totally negligible because it is proportional to neutrino mass-squared 
differences, the former is important because the mass of $S_3$ is 
comparable to $m_{W_R}$ but $S_{1,2}$ are essentially massless.

The largest exotic contribution to $\mu \to e e e$ actually comes from the 
effective $Z \mu \bar e$ vertex.  The reason is that in this model,
\begin{equation}
Z = (1-x)^{1 \over 2} W^0_L - \left( {x^2 \over 1-x} \right)^{1 \over 2} 
W^0_R - \left( {x-2x^2 \over 1-x} \right)^{1 \over 2} B,
\end{equation}
hence a new vertex $Z W_R^+ W_R^-$ appears, in analogy to $Z W_L^+ W_L^-$ 
of the standard model.  The calculation of the one-loop $Z \mu \bar e$ 
vertex is similar to that of $Z d \bar s$\cite{mapr1} in the standard 
model.  The result is $g_{Z \mu \bar e} Z^\lambda \bar e [\gamma_\lambda 
(1+\gamma_5)/2] \mu$, with
\begin{equation}
g_{Z \mu \bar e} = {e^3 U_{\mu 3} U_{e3} \over 16 \pi^2 x^{1 \over 2} 
(1-x)^{1 \over 2}} \left[ {r_3 \over 1-r_3} + {r_3^2 \ln r_3 \over (1-r_3)^2} 
\right],
\end{equation}
where
\begin{equation}
r_3 \equiv m_{S_3}^2/m_{W_R}^2 = \left( {1-x \over 1-2x} \right) \left( 
1 + {v_2^2 \over v_3^2} \right)^{-1}.
\end{equation}
Hence this vertex is not suppressed at all, and its contribution to the 
$\mu \to e e e$ decay amplitude is proportional to $1/m_Z^2$, so it is 
much larger than that of the box diagram from $W_R^\pm$ and $S_i$ exchange, 
which is proportional to $1/m_{W_R}^2$.

Connecting the $Z \mu \bar e$ vertex with the standard-model $Z e \bar e$ 
vertex, the decay branching fraction of $\mu \to e e e$ is then given by
\begin{equation}
B(\mu \to e e e) = 2x(1-x)(1-4x+12x^2)g^2_{Z \mu \bar e}/e^2.
\end{equation}
Since the present experimental upper limit\cite{pdg} of this is $1.0 \times 
10^{-12}$, the following constraint is obtained:
\begin{equation}
U_{\mu 3} U_{e 3} < 2.3 \times 10^{-3},
\end{equation}
where the $v_2^2/v_3^2$ term of Eq.~(23) has been neglected.

(C) Because of Eq.~(22), the rare decay $Z \to \mu^- e^+ + e^- \mu^+$ is 
also predicted.  However, because of Eq.~(25), its branching fraction is 
less than $8.6 \times 10^{-13}$ which is of course totally negligible. 
The analogous decays $Z \to \tau^- e^+ + e^- \tau^+$ and $Z \to \tau^- 
\mu^+ + \mu^- \tau^+$ are related to $\tau \to e e e$, $\tau \to e \mu \mu$, 
$\tau \to \mu e e$, and $\tau \to \mu \mu \mu$, with upper limits\cite{pdg} 
of order $10^{-5}$ and $10^{-6}$ on their branching fractions respectively. 
All are predicted in this model to have branching fractions of order 
$10^{-7}$ multiplied by $U^2_{\tau 3} U^2_{e3}$ or $U^2_{\tau 3} U^2_{\mu 3}$.

(D) The archetypal rare decay $\mu \to e \gamma$ is also predicted in this 
model, again through $W_R^\pm$ and $S_i$ exchange.  The one-loop diagrams 
are analogous to the usual ones in the standard model, resulting 
in a decay branching fraction
\begin{equation}
B(\mu \to e \gamma) = {3 \alpha \over 32 \pi} \left( {m_{W_L} \over m_{W_R}} 
\right)^4 U^2_{\mu 3} U^2_{e3} F^2(r_3),
\end{equation}
where the function $F$ is given by\cite{mapr2}
\begin{equation}
F(r_3) = {r_3(-1+5r_3+2r_3^2) \over (1-r_3)^3} + {6r_3^3 \ln r_3 \over 
(1-r_3)^4}.
\end{equation}
Using the most recent experimental upper bound\cite{brooks} of $1.2 \times 
10^{-11}$ on $B$, we obtain
\begin{equation}
U_{\mu 3} U_{e3}(m_{W_L}^2/m_{W_R}^2) < 3.8 \times 10^{-4}.
\end{equation}

(E) If the lightest exotic quark, call it $h_1$, is lighter than $W_R^\pm$, 
then its decay is predominantly given by
\begin{equation}
h_1 \to u_i e_j S_k.
\end{equation}
Since $S_{1,2}$ are light and undetected, this mimics the ordinary 
semileptonic decay of a heavy quark, but without any nonleptonic component. 

In conclusion, the skew E$_6$ left-right model proposed many years 
ago\cite{ma}, favored\cite{erla} by recent atomic physics\cite{bewi} 
and $Z$ resonance\cite{mnsw} data, has been shown to be a natural framework 
for 3 active and 2 sterile light neutrinos.  The constraints from low-energy 
data, as given by Eqs.~(20), (25), and (28), require at worst ($U_{\mu 3} = 
U_{e3}$) only that $m_{W_R} > 
442$ GeV, or equivalently $m_{S_3} (= m_{Z'}) > 528$ GeV.  Hence the new 
physics of this model is accessible to experimental verification in the 
not-so-distant future.

This work was supported in part by the U.~S.~Department of Energy under Grant 
No.~DE-FG03-94ER40837.

\bibliographystyle{unsrt}

\begin{thebibliography}{99}
\bibitem{sterile} A sterile neutrino is one which is a singlet under the 
standard-model gauge group.  However, it may have other interactions which 
are suppressed at low energies.
\end{thebibliography}

\end{document}